\documentclass{article}
\usepackage{spconf,amsmath,graphicx}
\usepackage{url}

\newcommand\blfootnote[1]{%
	\begingroup
	\renewcommand\thefootnote{}\footnote{#1}%
	\addtocounter{footnote}{-1}%
	\endgroup
}

\title{Towards Linking the Lakh and IMSLP Datasets}
\name{TJ Tsai}
\address{Harvey Mudd College, Claremont, CA}

\begin{document}
\ninept
\maketitle

\begin{abstract}
This paper investigates the problem of matching a MIDI file against a large database of piano sheet music images.  Previous sheet--audio and sheet--MIDI alignment approaches have primarily focused on a 1-to-1 alignment task, which is not a scalable solution for retrieval from large databases.  We propose a method for scalable cross-modal retrieval that might be used to link the Lakh MIDI dataset with IMSLP sheet music data.  Our approach is to modify a previously proposed feature representation called a symbolic bootleg score to be suitable for hashing.  On a database of 5,000 piano scores containing 55,000 individual sheet music images, our system achieves a mean reciprocal rank of $0.84$ and an average retrieval time of $25.4$ seconds.
\end{abstract}

\section{Introduction}
\label{sec:intro}

The goal of this paper is to propose and validate a method for linking two large-scale datasets in the music information retrieval community: the Lakh MIDI Dataset \footnote{\url{https://colinraffel.com/projects/lmd/}} \cite{raffel2016thesis} and the International Music Score Library Project (IMSLP) dataset.\footnote{\url{https://imslp.org}}  The Lakh dataset is a collection of 176,581 unique MIDI files that were scraped from publicly-available sources on the internet.  The IMSLP dataset contains nearly 500,000 sheet music scores representing 150,000 works and 18,000 composers.  Whereas IMSLP contains a rich set of metadata for each sheet music score, the Lakh dataset contains no organized metadata at all --- even the names of the files are simply their MD5 checksums.  Previous works \cite{raffel2015large}\cite{raffel2016optimizing}\cite{raffel2016pruning} have explored matching Lakh data to short audio preview recordings from the Million Song Dataset \cite{bertin2011million}.  To the best of our knowledge, this is the first attempt to link Lakh to a dataset of sheet music images.  This is a large-scale cross-modal retrieval problem.\blfootnote{
Citation information: DOI  10.1109/ICASSP40776.2020.9053815, Proceedings of the IEEE International Conference on Acoustics, Speech, and Signal Processing (ICASSP) 2020.

(c) 2020 IEEE.  Personal use of this material is permitted.  Permission from IEEE must be obtained for all other uses, in any current or future media, including reprinting/republishing this material for advertising or promotional purposes, creating new collective works, for resale or redistribution to servers or lists, or reuse of any copyrighted component of this work in other works.
}

Several previous works have investigated cross-modal alignment between sheet music images and audio.  Two general categories of approaches have been proposed.  The first approach is to convert the sheet music images to a symbolic representation using optical music recognition (OMR), to collapse the pitch information across octaves to get a chroma representation, and then to compare this representation to chroma features extracted from the audio.  This approach has been applied to synchronizing audio and sheet music \cite{DammFKMC08_MultimodalPresentationofMusic_ICMI}\cite{KurthMFCC07_AutomatedSynchronization_ISMIR}\cite{ThomasFMC12_LinkingSheetMusicAudio_DagstuhlFU}, identifying audio recordings that correspond to a given sheet music representation \cite{FremereyMKC08_AutomaticMapping_ISMIR}, and finding the corresponding audio segment given a short segment of sheet music \cite{FremereyCME09_SheetMusicID_ISMIR}.  The second approach is to convert both sheet music and audio into a learned feature space that directly encodes semantic similarity.  This has been done using convolutional neural networks combined with canonical correlation analysis \cite{dorfer2016towardsEnd}\cite{dorfer2018end}, pairwise ranking loss \cite{dorfer2017learning}\cite{dorfer2018tismir}, or some other suitable loss metric.  This approach has been explored in the context of online sheet music score following \cite{dorfer2016live}, sheet music retrieval given an audio query \cite{dorfer2016towards}\cite{dorfer2017learning}\cite{dorfer2018tismir}, and offline alignment of sheet music and audio \cite{dorfer2017learning}.  Dorfer et al. \cite{dorfer2018learningToListen} have also recently shown promising results formulating the score following problem as a reinforcement learning game.  See \cite{mueller2019cross} for a recent overview of work in this area.

Two recent works have explored cross-modal alignment between sheet music images and MIDI.  The first of these works \cite{tanprasert2019midisheet} takes the approach of converting MIDI into image pixel space, where note onsets are translated into floating rectangular notehead blobs placed appropriately on a blank image canvas containing the same staff line coordinate system as the sheet music.  This representation is called a pixel bootleg score.  The alignment can then be performed by directly comparing the similarity between columns of pixel values.  A related work explores an application in which a user would like to retrieve a passage of music from a MIDI file by taking a cell phone picture of a page of sheet music \cite{yang2019midipassage}.  This work proposes a feature representation called a symbolic bootleg score which encodes the position of noteheads relative to the staff lines.

The approaches described above are not viable solutions to the current task for one simple reason: they are not scalable.  These works have focused primarily on how to bridge the sheet--audio or sheet--MIDI modality gap within the context of a pairwise comparison.  This approach will not scale to a database as large as IMSLP.  The main challenge of the current task is to extend cross-modal alignment methods to large-scale retrieval.

Our approach to this problem is to modify the symbolic bootleg score features proposed in \cite{yang2019midipassage} to be suitable for hashing.  Even though these features were originally designed for use within a dynamic time warping framework, we show that they can be adapted to function effectively in a reverse-indexing scheme.

This paper has two main contributions.  First, we introduce a curated dataset for studying large-scale MIDI--sheet music retrieval.  This dataset contains 200 MIDI files and 5,000 piano sheet music scores containing 55,000 individual sheet music images.  Since the IMSLP dataset takes more than a month just to download, we introduce a much more manageable dataset to facilitate research on this topic.  Second, we propose a method based on modified symbolic bootleg score features which uses a reverse-indexing scheme.  Our system is able to achieve a mean reciprocal rank of $0.84$ with an average retrieval time of $25.4$ seconds.

\begin{figure}
	\includegraphics[width=\columnwidth]{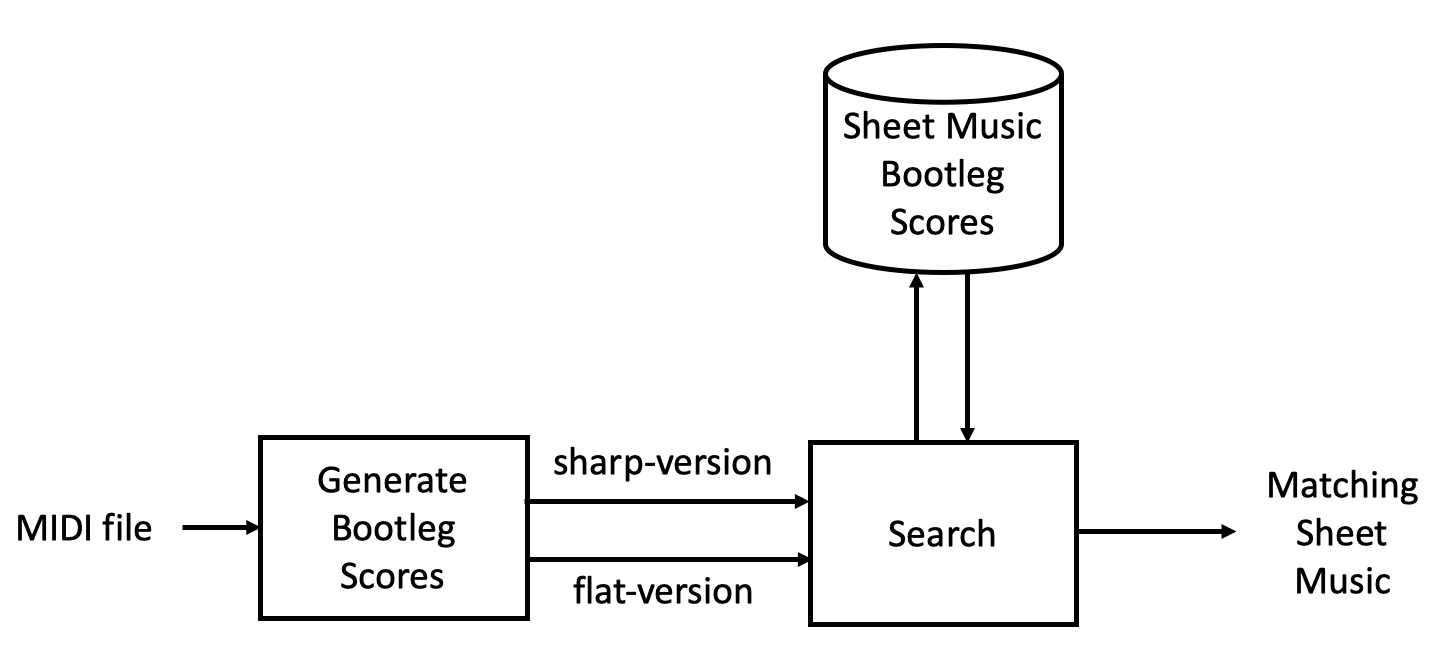}
	\caption{Architecture of proposed system.}
	\label{fig:systemOverview}
\end{figure}

\section{System Description}
\label{sec:systemDescr}

The overall architecture for our system is shown in Figure \ref{fig:systemOverview}.  In this work, we focus exclusively on solo piano music.  The MIDI file is processed to generate two different bootleg score representations (for reasons that will be discussed later), and these bootleg scores are used to search a database of sheet music bootleg scores.  There are three key components that are needed to construct this system: extracting bootleg score features from MIDI, extracting bootleg score features from sheet music images, and performing the database search.  Each of these three components will be described in the following three subsections.

\subsection{Extracting MIDI Bootleg Features}
\label{subsec:midiBootleg}

Extracting MIDI bootleg score features consists of three steps.  First, a list of note onsets and their onset times is generated.  Second, the note onsets are grouped into a sequence of note events, where a single note event consists of one or more note onsets that occur (approximately) simultaneously.  Third, the note events are projected onto a bootleg score in a manner described below.

\begin{figure}
	\includegraphics[width=\columnwidth]{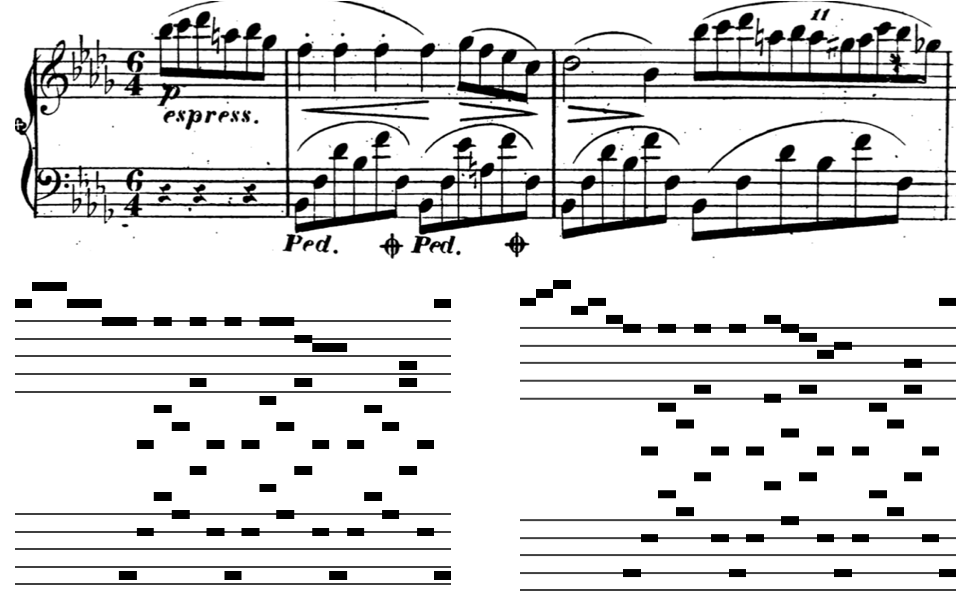}
	\caption{Comparison of the sharp-version bootleg score (bottom left) and the flat-version bootleg score (bottom right), along with the corresponding sheet music (top).  The staff lines are shown as a visual aid, but are not present in the actual feature representation.}
	\label{fig:sharpVsFlat}
\end{figure}

The bootleg score is a symbolic representation that describes the position of noteheads relative to staff lines in sheet music.  For example, if the MIDI note value 61 (C\#4) is played, the notehead could occur in four possible locations: as a C-sharp in the right hand (i.e. one ledger line below the upper staff in treble clef), as a D-flat in the right hand, as a C-sharp in the left hand (i.e. one ledger line above the lower staff in bass clef), or as a D-flat in the left hand.  (Though the note could occur in other positions due to clef changes, double sharps, or double flats, we ignore these since they are relatively uncommon.)  Each of these four possibilities corresponds to a different vertical location in a grand staff.  In the original formulation \cite{yang2019midipassage}, this ambiguity was handled by simply placing floating notehead blobs in all possible locations.  The problem with this approach is that the resulting representation will never match what is observed in the sheet music, since each note only occurs in one position.  Thus, we modify the original formulation by generating two separate bootleg scores: one which assumes that all black keys are sharps and the other which assumes that all black keys are flats.  As before, floating notehead blobs are placed in both the right hand and left hand locations for notes in the middle register.  The bootleg score is a binary matrix with dimensions $62 \times N$, where $N$ is the number of note events and $62$ is the number of different vertical locations in the grand staff.  The bootleg score spans from E3 to C8 in the right hand (34 positions) and from A0 to G4 in the left hand (28 positions).  In the same way that the C\#4 is projected onto each bootleg score as two floating notehead impulses, every note onset in the MIDI file can be projected onto the bootleg  score in a similar manner.  Figure \ref{fig:sharpVsFlat} shows the sharp- and flat-versions of the bootleg scores for a short segment of MIDI, along with the corresponding sheet music for reference.

\subsection{Extracting Sheet Music Bootleg Features}
\label{subsec:pdfBootleg}

\begin{figure}
	\includegraphics[width=\columnwidth]{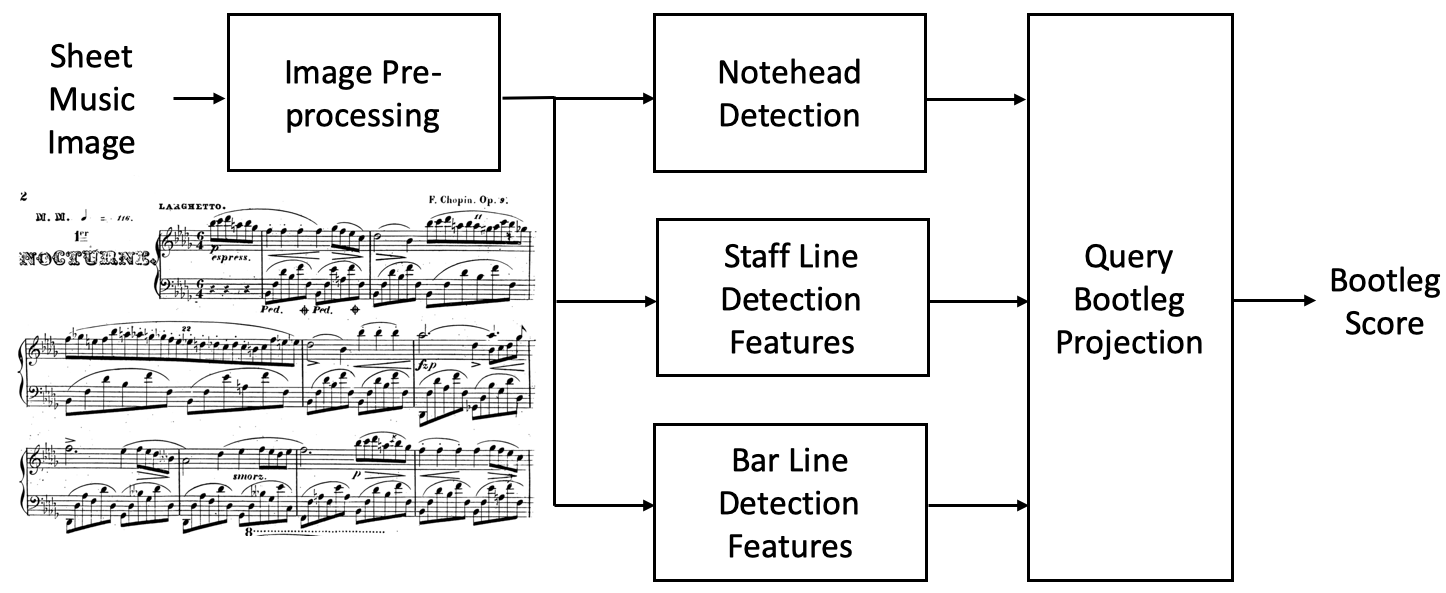}
	\caption{Extracting sheet music bootleg score features.  The bootleg scores from each sheet music image are concatenated to form a global bootleg score for the entire piece.}
	\label{fig:generatePdfBootleg}
\end{figure}

The process of extracting sheet music bootleg score features has five steps, as shown in Figure \ref{fig:generatePdfBootleg}.  Each of the five steps will be briefly described in the following five paragraphs.  For more details, the reader is referred to \cite{tsai2019using}.

The first step is to perform image pre-processing.  This includes converting to grayscale, removing background lighting by subtracting away a blurred version of the image, and performing interline normalization by computing the responses to a bank of differently sized comb filters and resizing the image according to the estimated staff line separation.

The second step is to detect filled noteheads in the image.  The feature extraction focuses only on filled noteheads because they generally occur much more frequently than half or whole notes, and because they are relatively easy to estimate with classical computer vision tools due to their simple geometrical shape (i.e. a circular blob).  We detect noteheads by: (a) filtering out other objects by eroding and dilating the image with a circular morphological filter, (b) applying a simple blob detector from OpenCV to estimate a template of the filled notehead blobs, (c) binarizing the eroded and dilated image to get a list of connected components, and (d) using the estimated template to select only the connected components that are of the expected size.  The result of this step is a list of detected notehead blobs.

The third step is to compute a set of features indicating the locations of staff lines.  The pre-processed image is  eroded and dilated with a short, fat morphological filter to remove everything except horizontal lines.  The resulting image is then convolved with a bank of differently sized vertical comb filters, where each comb filter corresponds to a different staff line spacing.  The result of this step is a feature tensor which indicates the size and location of staff lines in the image.

The fourth step is to compute a set of features indicating the vertical location of bar lines.  The pre-processed image is eroded and dilated with a tall, skinny morphological filter to remove everything except tall, vertical lines.  The bar line features are simply the sum of the pixel values in each row, where a large row sum indicates the presence of multiple barlines at that vertical pixel location.

The fifth step is to project the detected notehead blobs onto a bootleg score.  For each detected notehead blob, we infer the nearest staff line locations using the staff line features, and then estimate the notehead's vertical staff line location using simple linear interpolation.  We then group pairs of staves together using the bar line features, and use each pair of staves to generate a fragment of the bootleg score.  Finally, we perform one additional step that is not in the original design: we mirror left and right hand floating noteheads so that notes in the middle register appear in both the right and left hands.  The resulting bootleg score is a $62 \times M$ binary matrix, where $M$ indicates the number of estimated note events on the page.

One important characteristic about this feature extraction process is that there are no trainable weights -- only a set of about 40 hyperparameters.  The fact that there are no trainable weights makes this feature representation far less susceptible to overfitting.  Indeed, we are able to use the same feature extraction for scanned sheet music, even though the features were originally designed for a very different domain (cell phone pictures of sheet music).  In switching between these two domains, we kept all the hyperparameter values the same except for four: the minimum and maximum number of staves we expect to encounter in an image, and the minimum and maximum staff line spacing (before interline normalization).  The hyperparameters were already tuned for cell phone images in \cite{tsai2019using}, and we tuned these four hyperparameters for scanned sheet music on a set of 20 sheet music images, which were separate from the test data.  The tuning process took about 15 minutes of human time.

\begin{figure*}
	\includegraphics[width=\textwidth]{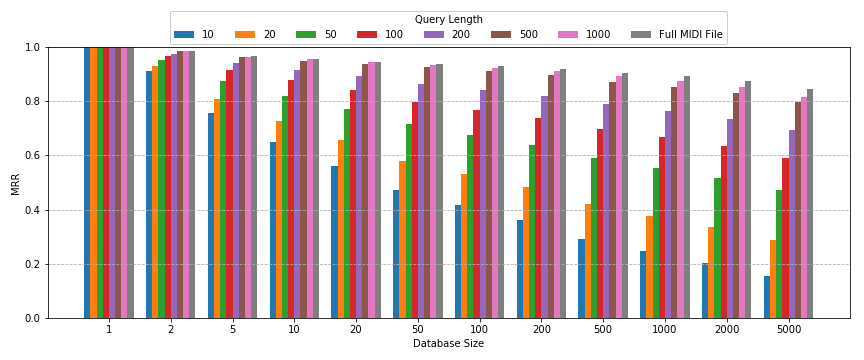}
	\caption{Comparison of system performance for various database sizes and MIDI query lengths.  Each bar shows the average performance across ten different simulations in which a database of size $N$ is randomly generated from the full set of $5024$ sheet music scores.}
	\label{fig:accuracy}
\end{figure*}

\subsection{Search}
\label{sec:search}

We perform the database search using the Shazam search method \cite{wang2003industrial}, where the bootleg score features are used in place of the peak pair audio fingerprints.  This method has two stages.  In the first stage, we compute bootleg score features on a database of sheet music images and construct a reverse index on the bootleg score columns, which we will refer to as bootleg fingerprints.  Note that each bootleg fingerprint consists of 62 bits, so it can be efficiently represented as a single 64-bit integer.  The reverse index is a mapping from a bootleg fingerprint value to a list of tuples indicating the piece and offset of each occurrence of the fingerprint value.  In the second stage, we compute bootleg features on the MIDI query and use the reverse index to construct a histogram of $(t_{ref} - t_{query})$ values for matching fingerprints, where $t_{ref}$ and $t_{query}$ are the offsets in the database bootleg score and query, respectively.  The maximum histogram bin count is used as a match score to sort the pieces in the database.

One additional change to the original formulation in \cite{wang2003industrial} was needed to make the system scalable.  The distribution of bootleg fingerprint values is Zipfian in shape, with an extremely sharp peak and a very long tail.  On the database of 5000 scores, there were about 216,000 unique fingerprint values, of which 120,000 only occurred a single time.  On the other hand, the most frequent bootleg fingerprint occurred 187,000 times.  Because some bootleg fingerprint values (particularly those that have only a single notehead) occur very frequently, these fingerprints are not very informative and require the search to process lots of spurious matches.  To address this issue, we replaced fingerprints that occurred more than $8000$ times in the database with a fingerprint triplet which encodes a sequence of three fingerprints.  This ensures that no fingerprint (either single or triple) occurs too frequently and spreads the distribution over a much wider number of unique fingerprint values (944,000 vs. 216,000).

\begin{table}
	\begin{center}
		\begin{tabular}{| l | c | c | c |}
			\hline
			System & DB & Pages & $T_{avg}$\\
			\hline
			RetinaNet \cite{lin2017focal} & - & $1$ & $11.7$s\\
			Sheet--Audio Align \cite{dorfer2018tismir} & - & $1$ & $17.5$s \\
			Faster R-CNN \cite{ren2015faster} & - & $1$ & $49.9$s\\
			DWD \cite{tuggener2018deepwatershed} & - & $1$ & $213.1$s\\			
			Bootleg-DTW \cite{yang2019midipassage} & - & $1$ & $0.90$s\\
			\hline
			RetinaNet \cite{lin2017focal} & $5$k & $55$k & $178$h\\
			Sheet--Audio Align \cite{dorfer2018tismir} & $5$k & $55$k & $266$h \\
			Faster R-CNN \cite{ren2015faster} & $5$k & $55$k & $759$h\\
			DWD \cite{tuggener2018deepwatershed} & $5$k & $55$k & $3240$h\\			
			Bootleg-DTW \cite{yang2019midipassage} & $5$k & $55$k & $13.7$h\\
			\hline
			Bootleg-Shazam & $5$k & $55k$ & $25.4$s \\
			\hline
		\end{tabular}
	\end{center}
	\caption{Comparison of average runtime.  The upper cell shows previously reported runtimes from \cite{tsai2019using} on a single page sheet--MIDI alignment task.  The middle cell shows the corresponding estimated average runtime per query of these systems on the proposed task.  The lower cell shows the average runtime of the proposed system.}
	\label{tab:runtime}
\end{table}

\section{Results}
\label{sec:setup}

The data was adopted from \cite{tsai2019using} and augmented with additional sheet music scores to enable a database search.  The original dataset contains 200 piano scores downloaded from IMSLP and 200 matching MIDI files.  We use the same 400-1600 train-test split as \cite{tsai2019using}.  Since our system has no trainable weights -- only hyperparameters -- we use most of the data for testing.  We augmented the database by adding $5024$ scores containing $54,733$ individual sheet music images.  We considered MIDI queries of various lengths by randomly sampling intervals of length $L$ from the MIDI bootleg scores.  We sampled each MIDI file $10$ times, resulting in a total of $1600$ queries for each simulation.

Since each query has exactly one true match in the database, we use mean reciprocal rank (MRR) as an evaluation metric.  MRR is calculated as $\frac{1}{M} \sum \frac{1}{R_i}$, where $R_i$ is the rank of the true matching score for the $i^{th}$ query and $M=1600$ is the total number of queries.  Note that MRR ranges from $0$ to $1$, where $1$ corresponds to perfect performance.

Figure \ref{fig:accuracy} shows system performance across two different factors: database size and MIDI query length.  Each group of bars corresponds to a different database size, and the individual bars within each group correspond to different MIDI query lengths.  Since generating a database of size $N$ requires randomly selecting $N$ out of the $5024$ sheet music scores, we generate $10$ different databases of the desired size and average the results from all $10$ simulations.  

There are three things to notice about the results in Figure \ref{fig:accuracy}.  First, the system performance scales reasonably well with database size.  For example, as the database size increases from 5 to 50 to 500 to 5000, the MRR drops from $.96$ to $.93$ to $.90$ to $.84$.  Second, the MIDI query length becomes more important as database size increases.  As the search problem becomes more challenging, longer queries are needed to reliably identify the matching score.  Third, queries of length 500 and 1000 achieve almost as good performance as using the full MIDI file (which for piano solo works is typically in the thousands of note events), even up to the full database size of $5000$ scores.  This is an important observation because the MIDI and sheet music may not have a single one-to-one correspondence if the sheet music has structural jumps such as repeats or D.S. al Fine.  This means that we can break the MIDI file into shorter segments and perform the search with each segment, which will allow for finding matches even in the presence of structural jumps.

Table \ref{tab:runtime} compares the runtime of the proposed system to previously proposed cross-modal alignment approaches.  These approaches include a CNN-based sheet--audio alignment approach \cite{dorfer2018tismir}, the previously proposed bootleg system \cite{yang2019midipassage}, and several variants of the bootleg system augmented with state-of-the-art music object detectors \cite{tuggener2018deepwatershed}\cite{lin2017focal}\cite{ren2015faster} trained on the DeepScores dataset \cite{tuggener2018deepscores}.  The upper cell shows previously reported average runtimes from \cite{tsai2019using} on a sheet--MIDI alignment task between a single sheet music image and an entire MIDI file.  The middle cell shows the estimated average runtime per query on the proposed task if these systems were applied in a pairwise manner to each element of the database.  Note that the fastest of the previously proposed approaches -- the bootleg approach \cite{tsai2019using} -- would take $13.7$ \textit{hours} per query.  Aligning the Lakh and IMSLP datasets using this approach would take an estimated $50,000$ years.  The slowest approach based on the Deep Watershed Detector \cite{tuggener2018deepwatershed} would take $11.9$ million years without parallelization.  Assuming a linear scaling with database size, our proposed approach would take $25$ years without parallelization.

\section{Conclusion}
\label{sec:conclusion}

We have proposed a method for identifying a MIDI file within a large database of piano sheet music images.  Our approach is to modify a previously proposed symbolic bootleg score representation to be suitable for use with reverse-indexing.  We evaluate our system on a database of $5000$ sheet music scores from IMSLP containing a total of $55,000$ sheet music images.  Our system achieves a mean reciprocal rank of $0.84$ and an average runtime per query of $25.4$ seconds.  Future work includes further optimizing the system, scaling the system to bigger database sizes, and expanding the approach to work with non-piano music.

\bibliographystyle{IEEEbib}
\bibliography{LinkingLakhIMSLP}

\end{document}